\documentclass[11pt,english]{article}
\usepackage{amsmath}
\usepackage{graphicx}
\usepackage{epsfig}
\usepackage[lofdepth,lotdepth]{subfig}
\title{Probing CP violation at LBNE with Reactor experiments}
\author{Debajyoti Dutta, Kalpana Bora\\
Department of Physics, Gauhati University\\
Assam, Guwahati}
\usepackage{textcomp}
\usepackage{subfig}
\makeatother
\usepackage{babel}
\begin{document}
\maketitle
\begin{abstract}

 In this work, we have explored the possibilities of improving CP violation discovery potential of newly planned Long-Baseline Neutrino Experiment (LBNE), U.S.A., by combining with data from reactors. The third mixing angle $\theta_{13}$ is now very precisely measured and this precise measurenent of $\theta_{13}$ helps in measurement of CP violation. Here, CP violation is studied with and without data from reactors. Impact of placing a ND is also studied. It is found that CPV discovery potential of LBNE with ND increases when combined with data from reactors. With a far detector of 35 kt, it is possible to obtain 5$\sigma$ sensitivity of CPV when run for 5 years in $\nu$ and 5 years in $\bar{\nu}$ mode. When hierarchy is assumed to be normal hierarchy, CPV sensitivity is maximum. CPV discovery is possible by combining 5 years neutino data from LBNE with 3 years anti-neutrino data from reactors. This study reveals that CPV can also be discovered at 5$\sigma$ cl in IH mode when appearance measurement of LBNE is combined with reactors.\\

\end{abstract}

\section{Introduction}
Precise measurement of third neutrino mixing angle has opened up the door for physicists to search for its implementation in discovering other unknown parameters like neutrino mass hierarchy, CP violation etc. Specially, confirmation of non-zero $\theta_{13}$ \cite{1,2,3} has provided the experimental physicists a way to search for CP violation. CP violation in the leptonic sector can lead to leptogenesis \cite{4} which will help to answer unsolved questions like baryon asymmetry of the universe, baryogenesis etc \cite{5}. Although, CP violation in the neutrino sector is not directly linked to that in charged leptonic sector or in leptogenesis, discovery of CP violation in neutrino sector may add to the study of leptogenesis \cite{6}.\\
By measuring $\theta_{13}$ precisely and accurately, the new generation reactor experiments have done their job. Now it is time for the long baseline (LBL) experiments to explore the other sensitive issues in neutrino sector like determining the octant, mass hierarchy, CP phase etc. Actually, LBL experiments (LBNE \cite{7,8}, NOvA \cite{9}, T2K \cite{10}, MINOS \cite{11}, LBNO \cite{12,13} etc), have the advantage due to their long baselines. In LBL experiments due to their long baseline, matter effect \cite{14,15} comes in to play as it has opposite signs in the probability expression for the hierarchies and hence, they can differentiate between normal mass hierarchy (NH, $m^2_{3}-m^2_{2}>0$) and inverted mass hierarchy (IH, $m^2_{3}-m^2_{2} < 0$). The term that arises due to matter effect, changes sign when we change the oscillation from neutrino to anti neutrino mode or vice-versa. So, in presence of matter effect, CP is entangled and hence there are two degenerate solutions corresponding to one experiment. One solution corresponds to true values of CP and the other corresponds to the entangled value of CP. This problem can be resolved by combining two experiments with different baselines \cite{16,17}. For LBNE, baseline is so chosen that the asymmetry between $\nu_\mu \rightarrow \nu_e$ versus $\bar{\nu}_\mu \rightarrow \bar{\nu}_e$ is larger than the CP violating effect of $\delta
_{cp}$, which means that LBNE alone can determine mass hierarchy as well as CPV phase \cite{7}. But CPV measurement is dependent on $\theta_{13}$ as in neutrino mixing matrix, they appear together as $\theta_{13}e^{-i\delta}$(or $\theta_{13}e^{i\delta}$). To measure CP violation, $\sin^22\theta_{13}$ must be greater than or equal to 0.01. So, present knowledge of $\theta_{13}$ plays an important role in determining CP violation.\\
In this work, we have explored the CPV sensitivity of LBNE in presence of reactor experiments. Earlier, leptonic CPV has been explored by combining reactor experiment with superbeam experiment in \cite{18}. The combined sensitivity of a 10 kt LArTPC in conjunction with NOvA and T2K to discover the existence of leptonic CP violation is available in \cite{19}. They have found that for nearly 60$\%$ of the true $\delta_{cp}$ values, it is possible to discover CPV phase at 95$\%$ cl for the combined experiments. In \cite{20} and \cite{21}, authors have studied CPV for various configuration of LBNE with atmospheric data. Leptonic CP violation is also studied for the newly proposed high-power neutrino superbeam experiment \cite{22}. It is possible to discover CP violation at 2$\sigma$ cl for nearly 50$\%$ of true $\delta_{cp}$ by combining LBNE10 with LBNO \cite{23} experiments.The possibility of determining neutrino mass hierarchy independent of $\delta_{cp}$ at bi-magic baseline is studied in \cite{24,25,26}. The possibility of observing CP violation in the lepton sector in the proposed INO experiment, using the beta beam from CERN is studied in \cite{27}. \\
  The aim of this work is to achieve the best possible CPV sensitivity for LBNE in the light of moderately large value of $\theta_{13}$. We have combined LBNE with three reactor experiments together, Daya Bay, Double Chooz and RENO, to explore the different possibilities to improve CP violation sensitivity of LBNE.  Although, CP violation in neutrino sector is accessible to only appearance measurements \cite{7}, we have combined both appearance as well as disappearance measurements of LBNE in both neutrino and anti-neutrino modes with the data from reactors to study CPV sensitivity for 10 kt and 35 kt far detector. We have individully added Daya Bay and RENO experiments with 35 kt LBNE to study CPV. The effect of adding near detector in CPV sensitivity is also studied. Prior on $\theta_{13}$ plays an important role in octant sensitivity \cite{28}. Here, we have presented our results on CPV sensitivity in presence of prior on $\theta_{13}$. The effect of combining 5 years of neutrino data from LBNE (35 kt FD with ND), with 3 years of anti-neutrino data from reactor experiments, on CPV sensitivity is also presented here. The contribution made by the appearance channels $\nu_\mu \rightarrow \nu_e$ and $\bar{\nu}_\mu \rightarrow \bar{\nu}_e$ towards the CPV discovery in presence of reactor data has also been explored in this work. \\
  
  We find that CPV discovery potential of LBNE with ND can be improved significantly by combining with reactor experiments. Specially, if the assumed true hierarchy is NH, then the combined analysis can noticeably change the sensitivity. It is also possible to study CPV sensitivity at LBNE by combining its 5 years neutrino data with 3 years data from reactor experiments at 3$\sigma$ cl. We observed that only the appearance measurement at LBNE can discover CPV with more than 5$\sigma$ cl if data from reactors are added. We find that 10 kt (35 kt) FD with ND at LBNE can discover 45$\%$ (39$\%$) of true $\delta_{cp}$ at 3$\sigma$ (5$\sigma$) cl when combined with reactor experiments. These are the novel results presented in this work.\\
 
 The paper is organized as followed: in section 2, a theoretical discussion on CP violation in neutrino sector is presented. How the LBNE and reactors combination can explore CPV is also discussed here. Since we are combining a specific sets of experiments, in section 3, the details of the experiments are provided. The $\chi^2$ analysis as well as the systematic and statistical information are included in section 4. Section 5 contains results and discussions. In section 6, concluding remarks are included. 
 
\section{CPV in Neutrino Oscillation}
The precise and moderately large value of $\theta_{13}$ from reactor experiments have confirmed the non zero value of CP violating phase $\delta$. While studying CP violation, CP-asymmetry term, $A_{CP}$, plays the major role. It is the ratio of probabilities and is expressed as \cite{7,29}
 \begin{equation} A_{CP} = \frac{P(\nu_e\rightarrow \nu_\mu) - P(\bar{\nu}_e\rightarrow \bar{\nu}_\mu)}{P({\nu_e}\rightarrow \nu_\mu) + P(\bar{\nu}_e\rightarrow \bar{\nu}_\mu)}\end{equation} 
The series expansion of $P(\nu_e\rightarrow \nu_\mu)$ oscillation probability in vacuum upto $\alpha^2$ order can be approximated as \cite{29,30,31} :
 \begin{equation}P(\nu_e\rightarrow \nu_\mu) \approx P_\circ + P_{\sin\delta} + P_{\cos\delta} + P_3\end{equation} where, 
 \begin{equation}P_\circ = \sin^2\theta_{23}\sin^22\theta_{13}\sin^2\hat{\Delta}\end{equation}
 \begin{equation}P_{\sin\delta} = \alpha\sin\delta\cos\theta_{13}\sin2\theta_{12}\sin2\theta_{13}\sin2\theta_{23}\sin^3\hat{\Delta}\end{equation}
 \begin{equation}P_{\cos\delta} = \alpha\cos\delta\cos\theta_{13}\sin2\theta_{12}\sin2\theta_{13}\sin2\theta_{23}\cos\hat{\Delta}\sin^2\hat{\Delta}
\end{equation}   \begin{equation}P_3 = \alpha\cos^2\theta_{23}\sin^22\theta_{12}\sin^2\hat{\Delta}\end{equation}\\
 Where, $\alpha=\frac{\Delta m^2_{21}}{\Delta m^2_{31}}, $ $\Delta m^2_{31}= \Delta$, $\Delta m^2_{21}=\alpha\Delta$, $\Delta m^2_{32}=(1-\alpha)\Delta$ and $\hat{\Delta}=\frac{\Delta L}{4E}$.  The term containing the CP phase, i. e. $P_{\cos\delta}$ and $P_{\sin\delta}$ are suppressed by mass hierarchy as they contain terms upto O($\alpha$). In vacuum, $A_{CP}$ is directly proportional to $\sin\delta$ and hence it measures intrinsic CP violation.\\ In matter, due to matter effect (MSW effect), a new parameter, $A=\sqrt{2}G_FN_e$, appears in the probability expression and hence above expression for the appearance probability also changes in matter. Abbreviating $\hat{A}=\frac{A}{\Delta}$, the appearance probability $P_{\nu_\mu(\bar{\nu}_\mu)\rightarrow\nu_e(\bar{\nu}_e)}$ for LBNE in leading order in $\alpha=\Delta m^2_{21}/\Delta m^2_{31}$  can be written as \cite{29,31}
 \begin{equation} P(\nu_e\rightarrow \nu_\mu \cong P({\nu_\mu\rightarrow \nu_e}) \approx P_\circ + P_{\sin\delta} + P_{\cos\delta} + P_3\end{equation} where, neglecting all sub-leading terms in $\theta_{13}$, we obtain:
 
\begin{equation} P_\circ = \sin^2\theta_{23}\frac{\sin^22\theta_{13}}{(\hat{A}-1)^2}\sin^2((\hat{A}-1)\hat{\Delta})\end{equation}
\begin{equation}P_{\sin\delta} = \frac{\alpha\sin\delta\cos\theta_{13}\sin2\theta_{12}\sin2\theta_{13}\sin2\theta_{23}}{\hat{A}(1-\hat{A})}\sin\hat{\Delta}\sin(\hat{A}\hat{\Delta})\sin((1-\hat{A})\hat{\Delta})\end{equation}
\begin{equation}P_{\cos\delta} =\frac{\alpha\cos\delta\cos\theta_{13}\sin2\theta_{12}\sin2\theta_{13}\sin2\theta_{23}}{\hat{A}(1-\hat{A})} \cos\hat{\Delta}\sin(\hat{A}\hat{\Delta})\sin((1-\hat{A})\hat{\Delta})\end{equation}
\begin{equation}P_3 = \frac{\alpha\cos^2\theta_{23}\sin^22\theta_{12}}{\hat{A}^2}\sin^2(\hat{A}\hat{\Delta})\end{equation}
 
As $\hat{\Delta}\rightarrow 0$ for small baselines, all these expression converges to their counterpart in the vacuum oscillation probability. In presence of matter, if $\hat{A}\hat{\Delta}\leq 1$, then under the condition of bi-maximal mixing, CP asymmetry term is given as $$A_{CP}\sim \frac{1}{E_\nu}$$  So at high energies, $A_{CP}E_{\nu}$ is constant in the energy spectrum and hence it can directly access the CPV phase $\delta$ for any values of $\sin^22\theta_{13}$ \cite{29}.\\ As reactor experiments measure $\theta_{13}$ precisely, so when  $P({\nu_\mu\rightarrow \nu_e})$ of LBNE is combined with reactor experiments, they measure $\sin\delta$ \cite{18}. Since precise measurement of $\sin\delta$ is very much dependent on precise measurement of $\theta_{13}$, so adding reactor data with LBNE improves its CPV sensitivity noticeable.

% $$P_{\nu_\mu(\bar{\nu_\mu})\rightarrow\nu_e(\bar{\nu_e})} = 4s^2_{23}s^2_{13}(\frac{\bigtriangleup_{31}}{B_\mp}) \sin^2(\frac{B_\mp}{2})\pm 8c_{12}s_{12}c_{23}s_{23}\frac{\bigtriangleup_{21}}{AL}\frac{\bigtriangleup_{31}}{B_\mp}\sin(\frac{AL}{2})\sin(\frac{B_\mp }{2})\cos(\delta\pm\frac{\bigtriangleup_{31}}{2})$$\\
 %$$+c^2_{23}\sin^22\theta_{12}(\frac{\bigtriangleup_{21}}{AL})^2\sin^2(\frac{AL}{2})$$\\
 
 %Where, $B_\pm\equiv \bigtriangleup_{31}\pm AL$ and
% $\bigtriangleup^2_{ij}\equiv\frac{\Delta^2_{ij}L}{E}$. $A=\sqrt{2}G_FN_e$ is the refractive index of matter, $G_F$ is Fermi constant and $N_e$ is the electron no density of Earth. We can write this express of oscillation probability as:  
 
%$$P_{\nu_\mu(\bar{\nu_\mu})\rightarrow\nu_e(\bar{\nu_e})} = X_\pm s^2_{13} + Y_\pm s_{13}\cos(\delta\pm \frac{\bigtriangleup_{31}}{2}) + P_\odot$$, 
%where,  $X_\mp = 4s^2_{23}(\frac{\bigtriangleup_{31}}{B_\mp}) \sin^2(\frac{B_\mp}{2}) $, 
 %       $Y_\mp =\mp 8c_{12}s_{12}c_{23}s_{23}\frac{\bigtriangleup_{21}}{AL}\frac{\bigtriangleup_{31}}{B_\mp}\sin(\frac{AL}{2})\sin(\frac{B_\mp }{2})$ \\and 
 %       $$P_\odot = c^2_{23}\sin^22\theta_{12}(\frac{\bigtriangleup_{21}}{AL})^2\sin^2(\frac{AL}{2}) $$
%At oscillation maximum, $\Delta_{13}=\pi$, and then, the term$$(\cos\delta\pm\frac{\Delta_{13}}{2})=\mp\sin\delta$$, which is the CP violsting term in the probability expression. So for oscillation maximum, we can calculate $\sin\delta$ from:
%$$\sin\delta =\frac{P_{\nu_\mu(\bar{\nu_\mu})\rightarrow\nu_e(\bar{\nu_e})}-P_\odot- X_\pm s^2_{13}}{\mp Y_\pm s_{13}} $$
\section{Experimental Specifications}
In this section, we provide some details of the experimental set ups studied in this work. The informations are taken from respective literatures available so far.\\
LBNE \cite{7,8} is a world class accelerator facility planning to built in US to detect neutrino oscillation from Fermilab to Homestake mine where a non magnetized 10 kt Liquid Argon Time Projection Chamber (LArTPC) will detect neutrinos. The primary goal of LBNE is to explore the unknown sector of neutrino physics. The key strength of LBNE is its baseline which is 1300 km long and is sensitive to matter effect. Due to this sensitivity, it can alone resolve neutrino mass hierarchy with more than 3$\sigma$ precision. Matter effect induced asymmetry between $\nu_\mu\rightarrow \nu_e$ and $\bar{\nu}_\mu\rightarrow\bar{\nu}_e$ make it possible to measure both CPV and mass hierarchy within the same experimental set up. In the 1st phase of the experiment, it will run for 10 years with 5 years in $\nu$ and 5 years in $\bar{\nu}$ mode with 10 kt detector and 700 kw of power. Near detector will be employed for precision measurement of oscillation parameters. Use of ND will reduce the background systematic uncertainties. \\
 
 Double Chooz \cite{3,32}, Daya Bay \cite{1,33} and RENO \cite{2} are the three new generation reactor experiments with two detector set up which have proved their capability by precisely measuring the third neutrino mixing angle $\theta_{13}$. They are designed to detect $\bar{\nu}_e$ produced in inverse beta decay reaction. \begin{equation}
\bar{\nu}_{e}+p\rightarrow e^{+}+n\end{equation}
The four isotopes - $^{235}U$, $^{239}Pu$, $^{241}Pu$ and $^{238}U$, produce anti-neutrino flux at their respective reactor cores through $\beta$ decay reaction and then produced neutrinos are carried to the detector facility. The detail specifications of these three reactor experiments are tabulated below.
\begin{table}[!h]
\begin{center}
\begin{tabular}{|l|c|c|r|}
\hline 
Name of Exp  & Double Chooz  & Daya Bay  & RENO \tabularnewline
\hline 
Location  & France  & China  & Korea \tabularnewline
\hline 
No of Reactor cores  & 2  & 6  & 6 \tabularnewline
\hline 
Total Power(GW$_{th}$)  & 8.7  & 17.4  & 16.4 \tabularnewline
\hline 
Baselines- near/Far(m)  & 410/1067  & 360(500)/1985(1615)  & 292/1380 \tabularnewline
\hline 
Target mass(ND/FD (tons))  & 10/10  & 40 $\times$ 2/10  & 16.1/16.1 \tabularnewline
\hline 
Exposure(years)  & 3  & 3  & 3 \tabularnewline
\hline 
Resolution($\%$)  & 12  & 12  & 12 \tabularnewline
\hline
\end{tabular}
\caption{Detail specifications of the three reactor experiments}
\par\end{center}
\end{table}

Both Double Chooz and RENO have one near and one far detector in their detector site. But Daya Bay experimental site is little complicated and can be best understood from their collaboration report. Initially, Daya Bay had four reactor cores, two at Daya Bay NPP and other two at Ling Ao NPP. Later, two new cores at Ling Ao II were added. In this work, we have considered 3 years $\bar{\nu}$ data from the reactor experiments.
\section{Systematic Uncertainties and $\chi^2$ Analysis}
Systematic uncertainties play an important role in precision measurements. Reactor measurements are also dependent on systematic uncertainties like reactor related, detector related, and back-
ground related. But, using two identical detectors, the detector related systematic uncertainties
could be minimized and it also helps in reducing reactor related uncertainties. Considering the partial cancellation of errors due to the presence of near-far detectors, We have used a reduced set of errors (table 2) from their respective literatures \cite{34,35,36}. We have used GLoBES software \cite{37,38} for simulation. \\
\begin{table}[!t]
\begin{center}
\begin{tabular}{|l|l|c|c|}
\hline 
Name of Exp & RENO & Double-Chooz  & Daya Bay   \tabularnewline
\hline 
Reactor correlated error($\%$)  & 0.5 & 0.06  & 0.12  \tabularnewline
\hline 
Detector normalisation error($\%$) & 0.5  & 0.06  & 0.12  \tabularnewline
\hline 
Scaling or calibration error($\%$) & 0.1  & 0.5  & 0.5  \tabularnewline
\hline 
Overall normalization error($\%$) & 0.5  & 0.5  & 0.5  \tabularnewline
\hline 
bin to bin error($\%$) & 0.6  & 1.0  & 0.5  \tabularnewline
\hline
Spectrum shape related error($\%$) & 2.0  & 2.0  & 2.0  \tabularnewline
\hline
\end{tabular}
\caption{set of errors used in our calculation}
\par\end{center}
\end{table}
For LBNE, used energy range is 1-10 GeV while for reactors, it is 1.8-8 MeV. Event reconstruction efficiency is 85$\%$ for LBNE. All other informations like resolution, signal and background efficiencies are taken from \cite{7}. LBNE is also planning to employ near detector with 35 kt FD to reduce background systematic uncertainties. We have presented our results with and without ND both for 10 kt and 35 kt FD. The near detector informations are taken from \cite{20,21,39}.\\
The $\chi^2$ function, used for reactor experiments is:

$$ \chi^{2}=\underset{i=1}{\sum^{\#bins}}\underset{detector=ND,FD}{\sum}\frac{(O_{detector,i}-(1+a_{Reactor}+a_{detector})T_{detector,i})^{2}}{O_{detector,i}}$$
 \begin{equation}
 +\frac{a_{Reactor}^{2}}{\sigma_{Reactor}^{2}}+\frac{a_{ND}^{2}}{\sigma_{ND}^{2}}+\frac{a_{FD}^{2}}{\sigma_{FD}^{2}}
 \end{equation}
where, $a_{Reactor}, a_{FD},a_{ND}$ are reactor flux and detector mass related uncertainties,  $\sigma_{Reactor},$ $\sigma_{FD},\sigma_{ND}$ are respective standard deviations, $O_{ND,i}$,O$_{FD,i}$ are the event rates for the true values of oscillation parameters and T$_{detector,i}$ is the expected event rates for the i-th bin for the test values.\\
In this work, we have used minimised $\chi^2$ to study CPV sensitivity of LBNE. To calculate minimised $\chi^2$, we have marginalised the test parameters over $\delta_{cp} \in [-\pi, \pi]$, $\theta_{13} \in [3^0, 11^0]$, $\theta_{23} \in [36^0, 45^0]$ and $|\bigtriangleup m^2_{31}| \in [2.19, 2.62]\times 10^{-3} eV^2$ range. Since CP is conserved at $0^\circ$ and $180^\circ$, so excluding these two values, we search for all other possible true values of $\delta_{cp}$ for LBNE and combined set of experiments at different cl, for an assumed hierarchy. In equation 7, the terms $P_{\sin\delta}$ and $P_{\cos\delta}$, also known as CP violating and conserving term, change with $\sin^22\theta_{23}$, while the 1st term, $P_\circ$, is proportional to $\sin^2\theta_{23}$. So, $P_{\sin\delta}$ and $P_{\cos\delta}$ depend on the true value of $\theta_{23}$, but $P_\circ$ depends on both magnitude and octant of $\theta_{23}$. Again, CP asymmetry also changes with octants for a given hierarchy (fig.1). So in this work, we have studied CPV sensitivity for the true values of $\theta_{23}$ in both the octants. In our view, this is important, as so far the octant has not been fixed. We have 1st calculated the minimized $\chi^2$ for LBNE only and then combined the three reactors with LBNE to calculate the minimized $\chi^2$ of the combined system. i.e.
         \begin{equation}\chi^2_{min} = \chi^2_{LBNE}+\chi^2_{Reactors}\end{equation}
   If the calculated $\chi^2_{min}$ is greater than 4.0, 9.0 and 25.0 for some fraction of true values of $\delta_{cp}$ for an assumed true hierarchy and octant, CP violation discovery is then possible for that fraction of $\delta_{cp}$ at 2$\sigma$, 3$\sigma$ and 5$\sigma$ cl.
Effect of adding prior on $\sin^22\theta_{13}$ in CP violation sensitivity is also studied. The prior is defined as:
\begin{equation}\chi^2_{prior} =( \frac{\sin^22\theta^{true}_{13}-\sin^22\theta_{13}}{\sigma(\sin^22\theta_{13})})^2\end{equation}
Prior on $\sin^22\theta_{13}$ with 1$\sigma$ error range is: $\sigma_{\sin^22\theta_{13}}$=0.01 \cite{1,2}. In presence of prior, $\chi^2$ minimum is calculated as:
   \begin{equation} \chi^2_{total}=min(\chi^2_{LBNE}+\chi^2_{prior})\end{equation}
We have added prior on $\sin^22\theta_{13}$ to LBNE only, assuming the present knowledge of $\sin^22\theta_{13}$ from reactor experiments.\\
\begin{figure}[ph]
\centerline{\includegraphics[width=2.5in]{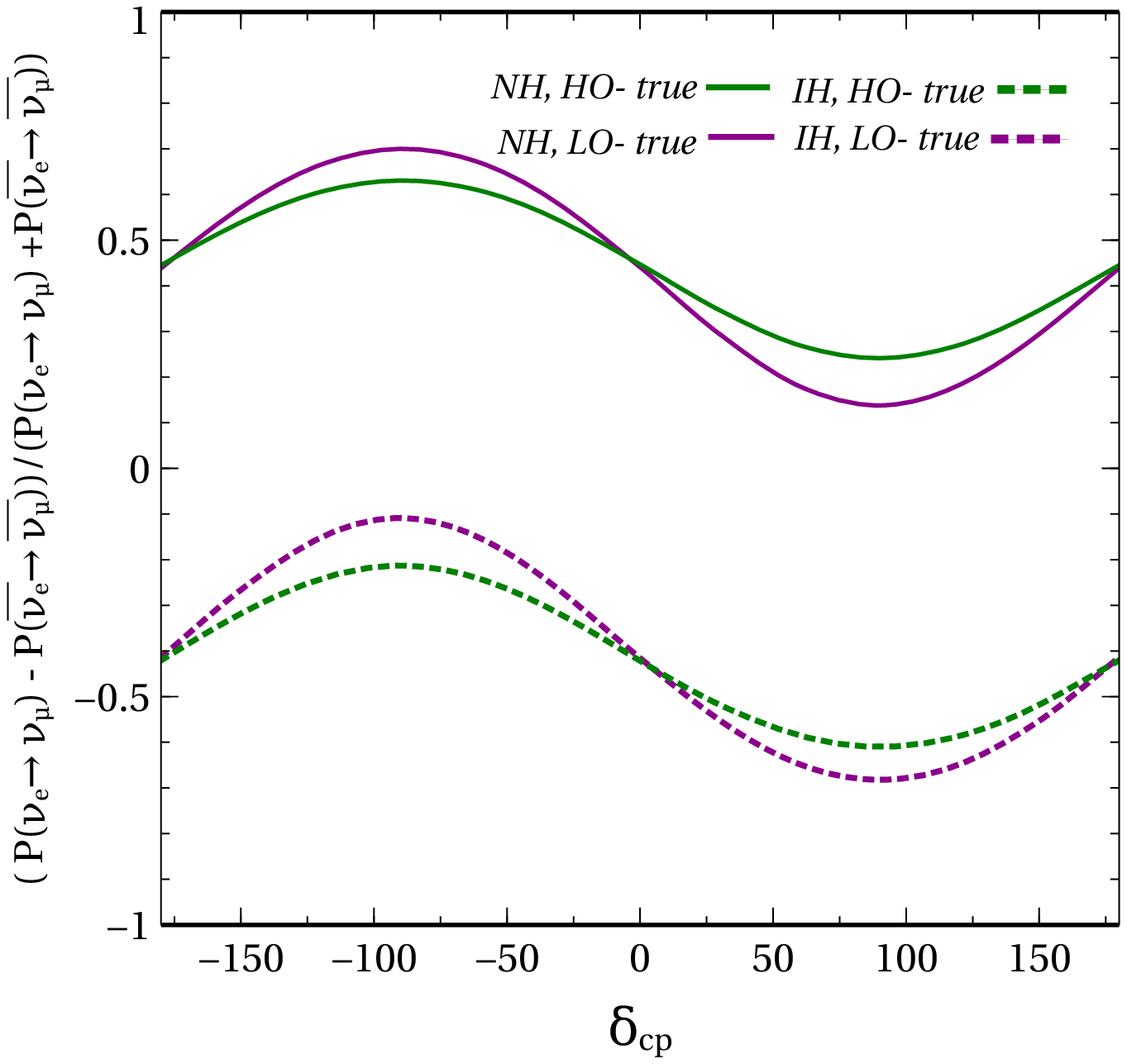}}
\vspace*{8pt}
\caption{ CP asymmetry vs $\delta_{cp}$ at LBNE, at 1st oscillation maxima for both the hierarchies. Green and darkmagenta solid (dotted) lines are for NH (IH) with HO and LO as the true octant respectively. \protect\label{fig1}}
\end{figure}

 \section{Results and discussions}
In this section, we present the results of this work. For the numerical calculation, we have used the full form of neutrino oscillation probability in matter. In fig.1, we have shown the variation of CP asymmetry term, $A_{CP}$, with $\delta_{cp}$ for both the hierarchies at LBNE. Assuming a true hierarchy, we have studied the asymmetry variation in both the octants. From fig.1, we find that CPV sensitivity at LBNE is
more in LO and in IH. In fig.2, we have shown the CPV sensitivity or possibilities of excluding the CP conserving phases for LBNE as well as the combined configurations under study. To generate these results, we have considered the global best fit values of neutrino oscillation parameters as the true values \cite{40}. Recent global data \cite{41} prefers the higher octant as the true octant, but they \cite{40,41} are agree with non- maximal $\theta_{23}$. Hence $\theta_{23}$ has two degenerate solutions corresponding to HO and LO. With new global data, we have generated one CP violation sensitivity plot for 35 kt FD (LBNE with ND), considering the LO as the true octant and NH as the true hierarchy (black solid plot in fig.3). The CP violation sensitivity achieved for this case is found to differ negligibly from that with the old global data \cite{40}. So, use of new global data will not remarkably change the results presented here. \\
 \begin{figure}[!t]
\begin{centering}
\begin{tabular}{cc}
\includegraphics[width=2.5in]{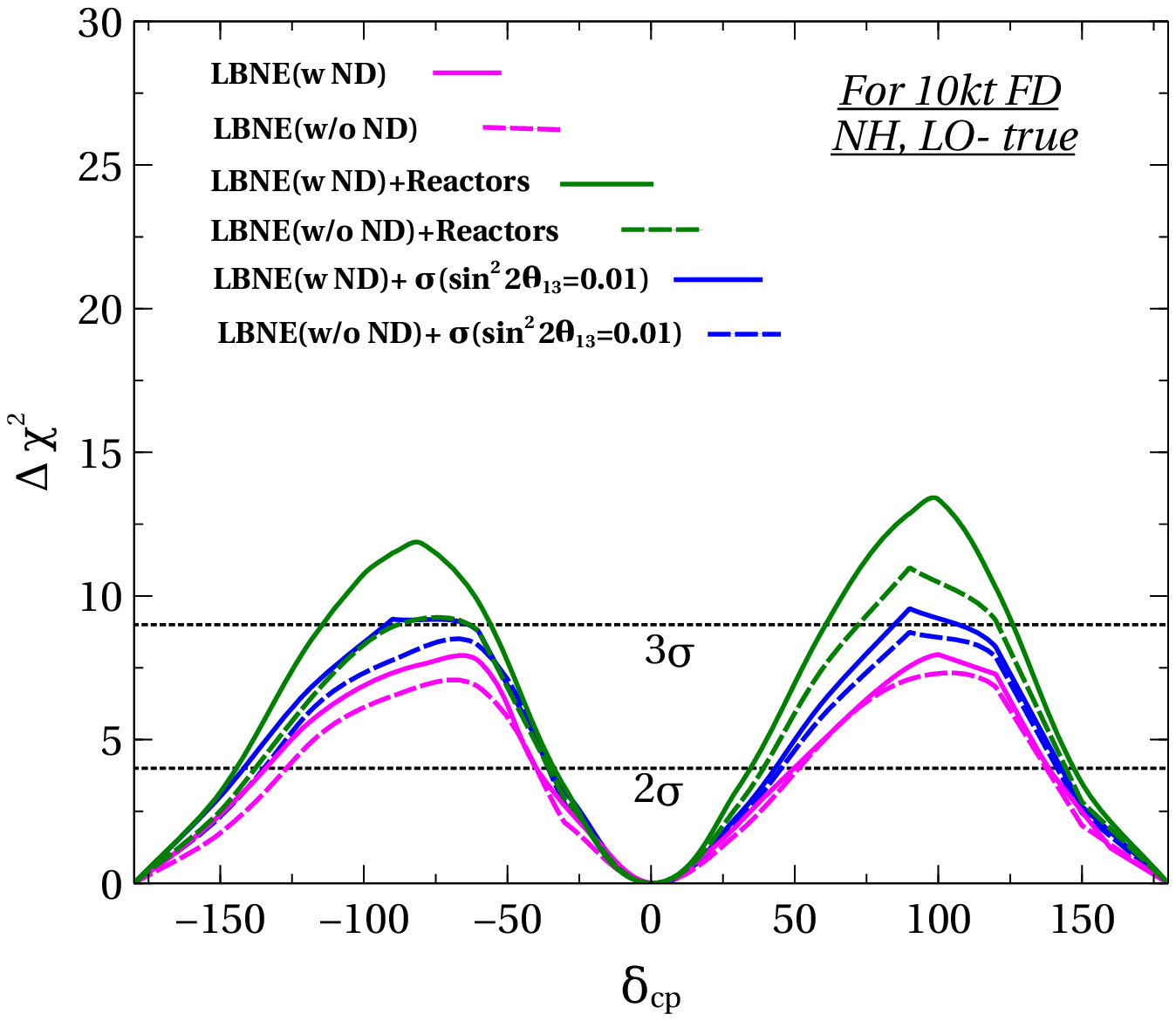}
\includegraphics[width=2.5in]{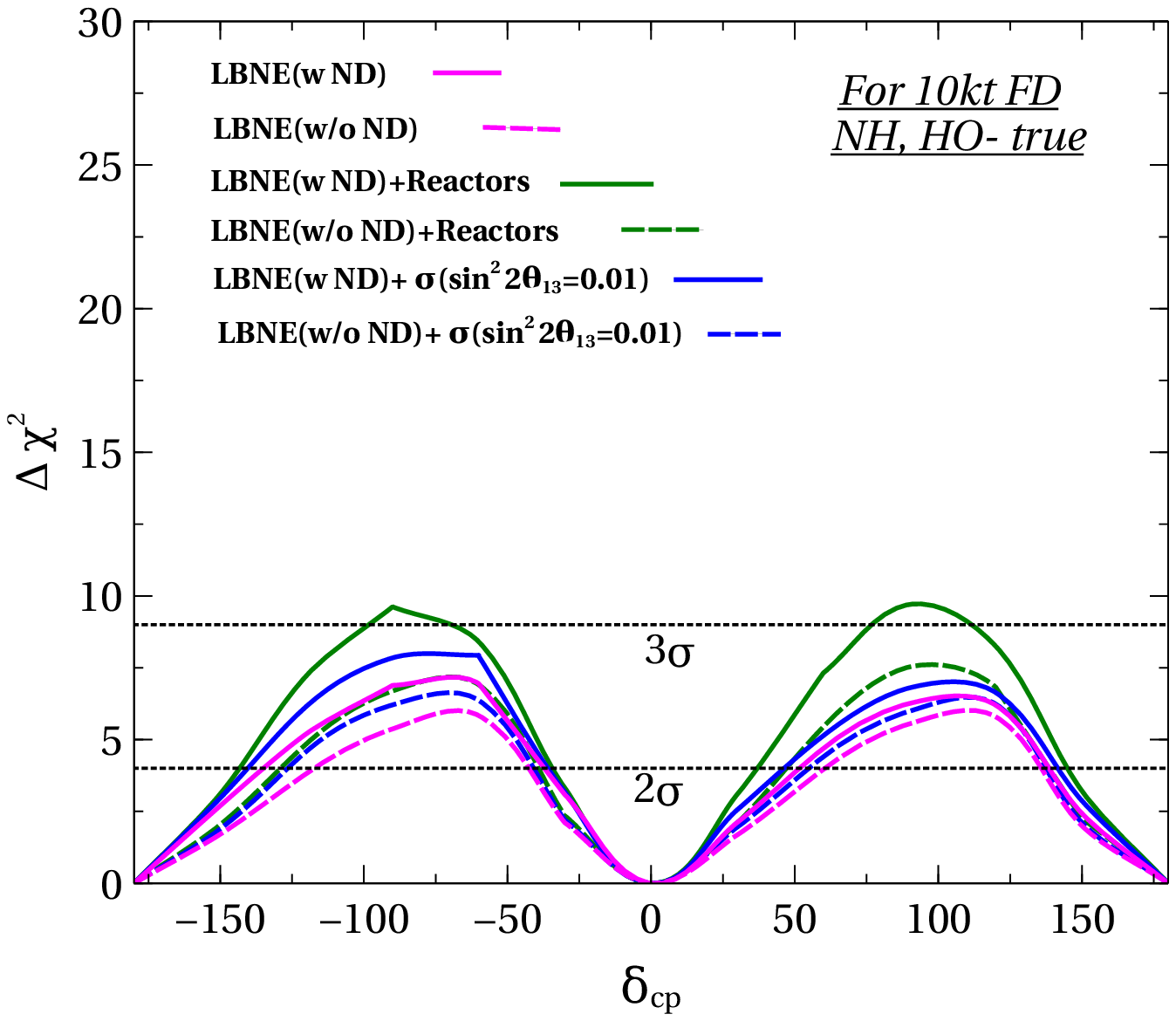}\\
\includegraphics[width=2.5in]{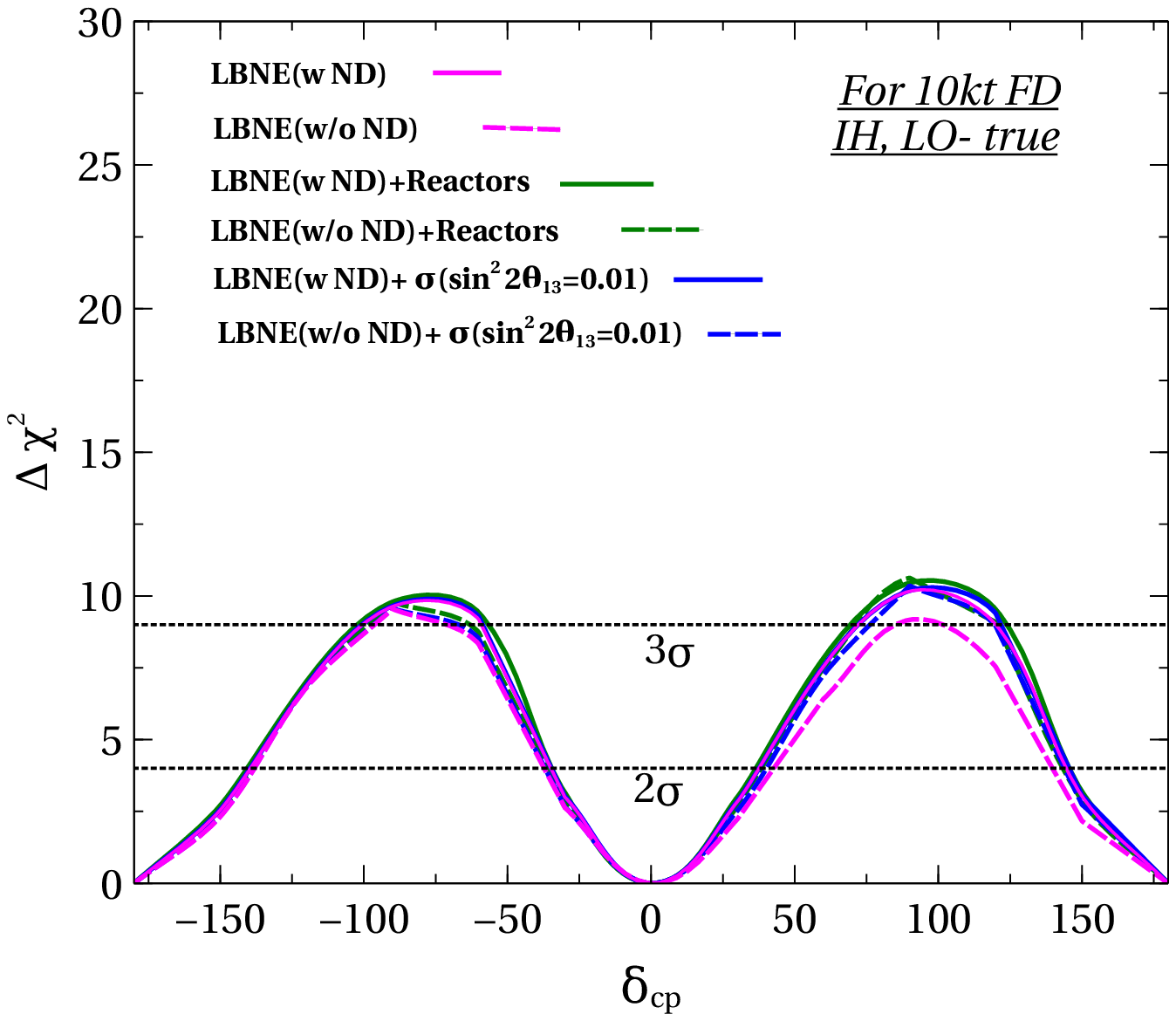}
\includegraphics[width=2.5in]{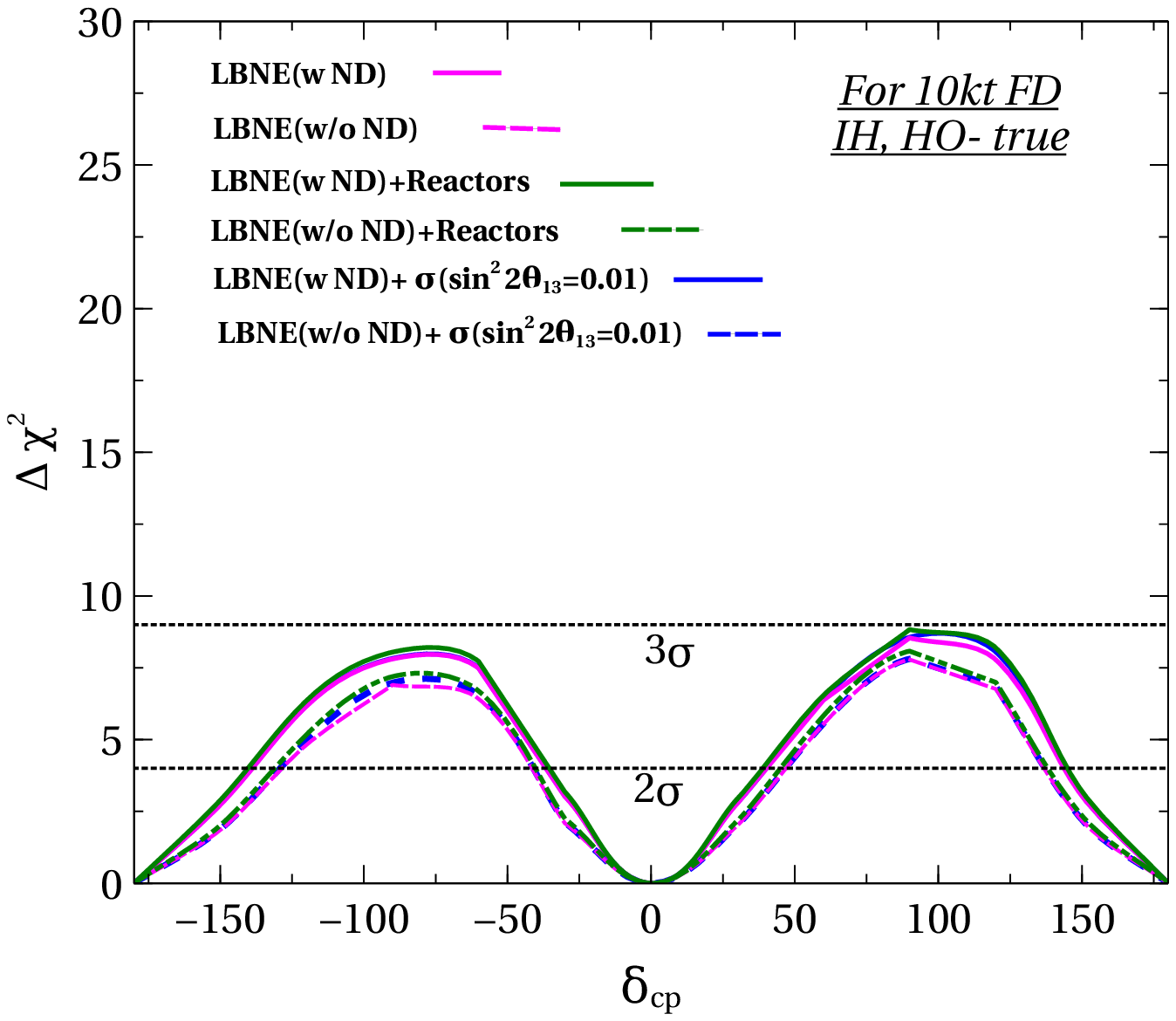}
\end{tabular}
\par\end{centering}
\caption{ CPV sensitivity plots for LBNE 10kt FD. In upper panel(lower panel), true hierarchy is assumed to be NH(IH). The plots are shown for 10 kt FD, with and without ND. We have presented CPV sensitivity for following possible combinations- LBNE only (magenta) (W and W/o ND), LBNE with prior (blue) (W and W/o ND) and LBNE with reactors (green) (W and W/o ND). The value of prior on $\sigma_{\sin^22\theta_{13}} = 0.01$. The solid plots are with ND while the dashed plots are without ND. Three reactor experiments DC, RENO and DB are added together with LBNE to examine the CPV sensitivity (Green plot). }
\end{figure}
\begin{figure}[!h]
\begin{centering}
\begin{tabular}{cc}
\includegraphics[width=2.5in]{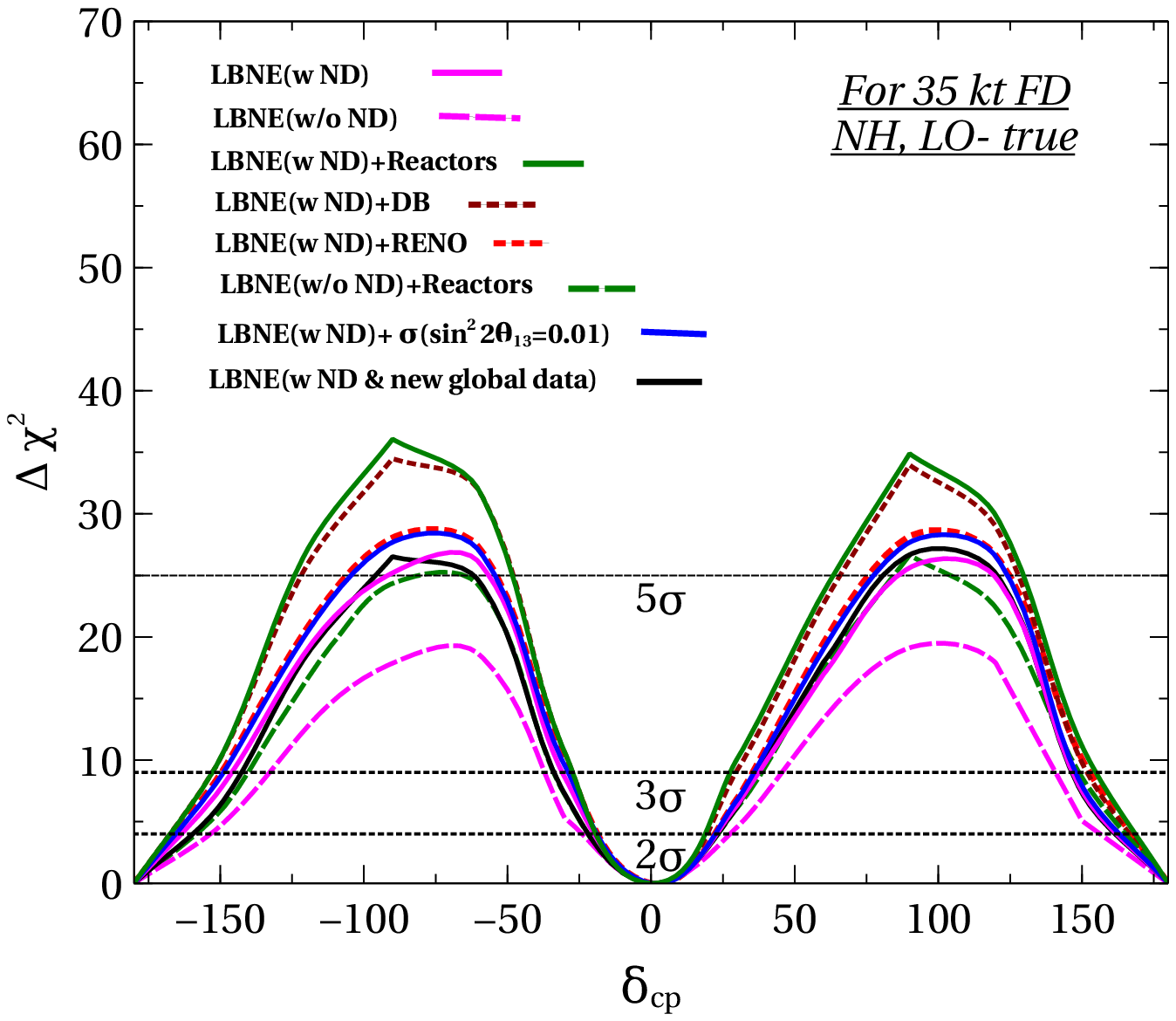}
\includegraphics[width=2.5in]{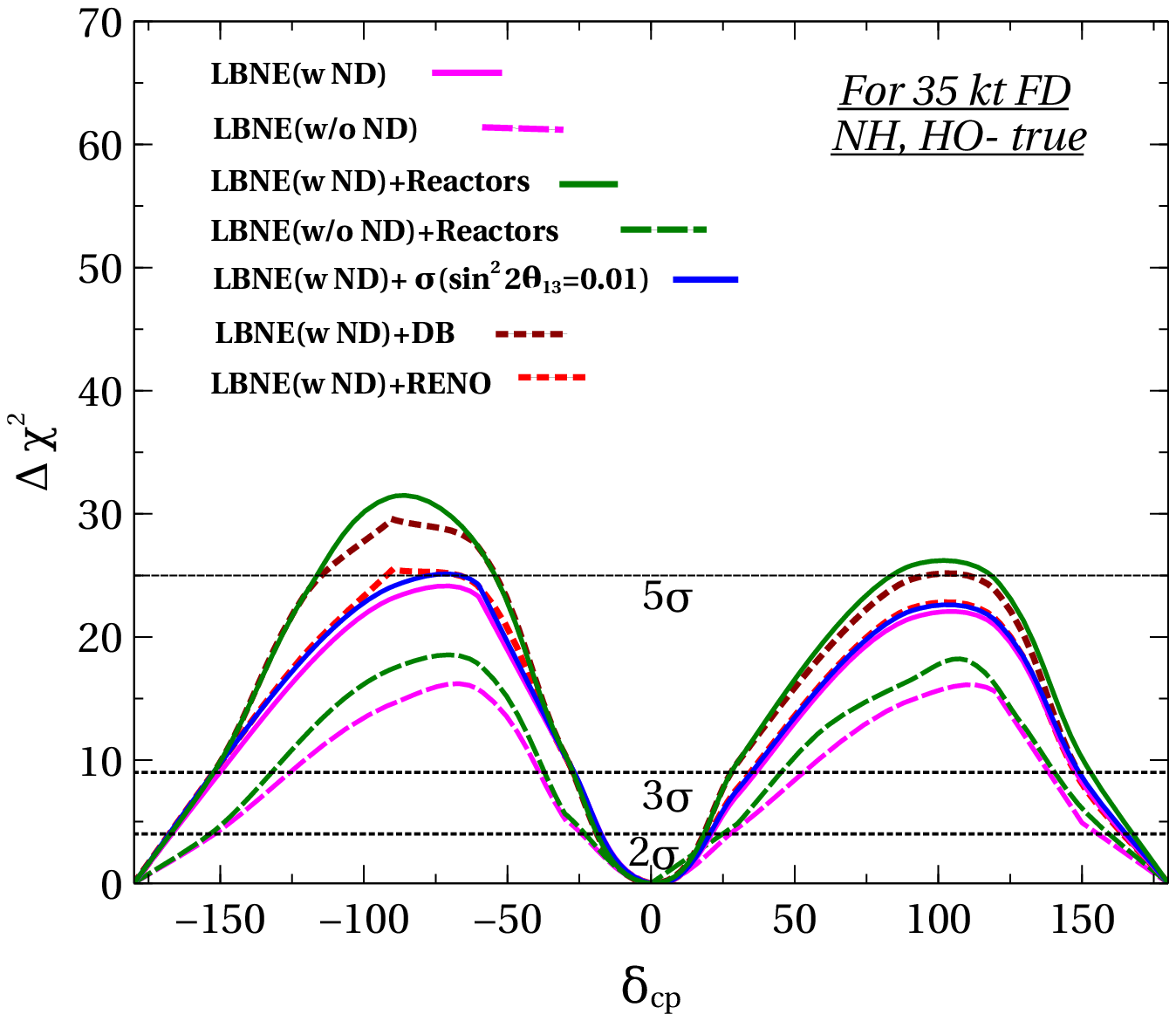}\\
\includegraphics[width=2.5in]{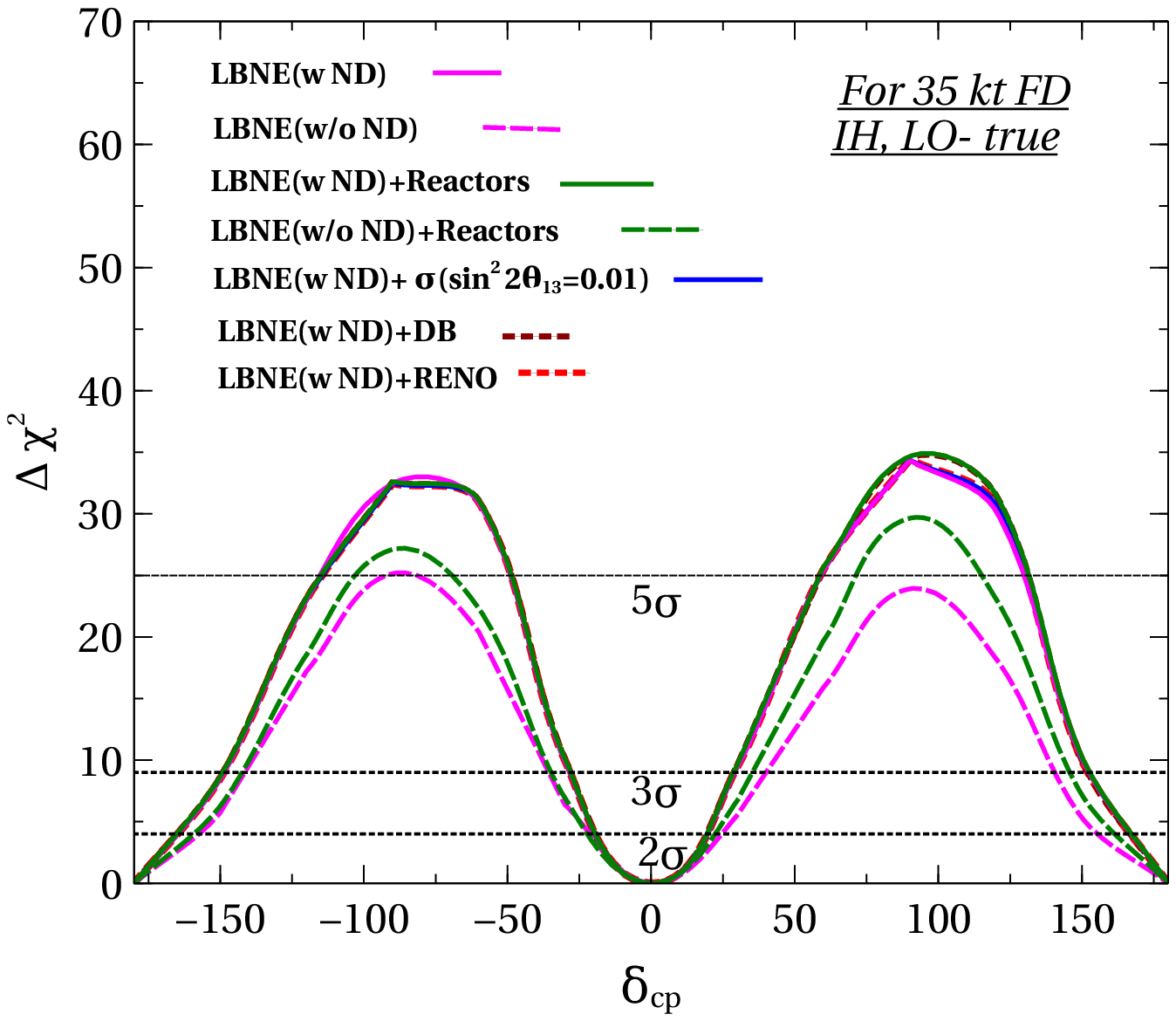}
\includegraphics[width=2.5in]{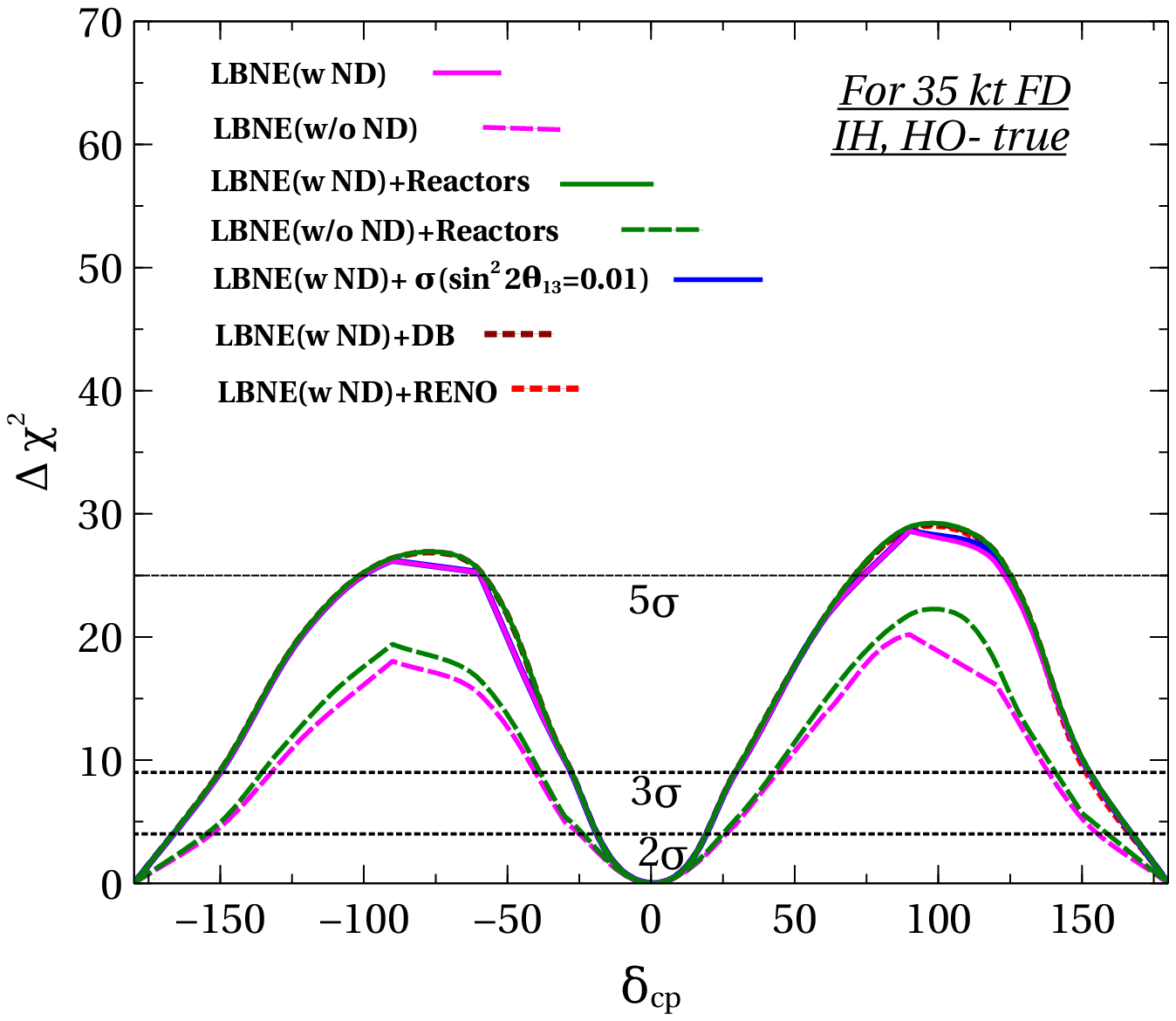}
\end{tabular}
\par\end{centering}
\caption{ CPV sensitivity plots for LBNE 35kt FD. In upper panel(lower panel), true hierarchy is assumed to be NH(IH). The plots are shown for 35 kt FD, with and without ND. We have presented CPV sensitivity for following possible combinations- LBNE only (magenta) (W and W/o ND), LBNE with prior (blue) (W and W/o ND) and LBNE with reactors (green) (W and W/o ND). The value of prior on $\sigma_{\sin^22\theta_{13}} = 0.01$. The solid plots are with ND while the dashed plots are without ND. Three reactor experiments DC, RENO and DB are added together with LBNE to examine the CPV sensitivity (Green plot). Individual contribution of RENO and Daya Bay towards CPV discovery is also explored with 35 kt FD (w ND) at LBNE. Plot for LBNE with prior (without ND) is not included as its effect is very negligible.}
\end{figure}
In fig.2 and fig.3, we have shown CP violation sensitivity plots for 10 kt and 35 kt far detectors at LBNE (with and without ND). Here, we have combined both $\nu_e$ appearance and $\nu_{\mu}$ disappearance  channel in neutrino and anti-neutrino mode. CPV sensitivity of above is then compared with that of LBNE in conjunction with reactor experiments. Effect of adding prior on $sin^22\theta_{13} = 0.01$ with LBNE is also presented. From fig.2 and fig.3, we make the following observations:
\begin{itemize}
\item From fig.2 and fig.3, it is observed that CP violation sensitivity is higher in LO for any assumed true hierarchy which is also confirmed by the asymmetry plot in fig.1.
\item In presence of ND, CPV sensitivity increases. Although this increase in CPV sensitivity of 10 kt FD is not so significant, but for 35 kt FD, ND information remarkably improves the CPV sensitivity.
\item CPV sensitivity increases as the mass of the FD increases. For 10 kt FD, 3$\sigma$ CPV discovery is possible for LBNE if true hierarchy and true octant are IH and LO respectively. For the same hierarchy and octant, with 35 kt FD, LBNE with ND alone can discover CP violation sensitivity at 5$\sigma$ cl. For this combination of hierarchy and octant, reactors combined LBNE produce similar result with LBNE alone. Presence of ND enhances the CPV sensitivity of 35 kt FD.
\item For 10 kt FD with ND, reactors combined LBNE can discover 0.45 fraction of $\delta_{cp}$ which corresponds to CP violation at 3$\sigma$ cl for (NH, LO-true) combination. For (IH, LO-true) combination, 0.27 fraction of $\delta_{cp}$ contributes to CPV at 3$\sigma$ cl for the same experimental set up. 
\item For 10 kt FD with ND, it is found that 0.16 fraction of CP violating $\delta$ can be discovered at 3$\sigma$ cl for (NH, HO-true) combination in presence of data from reactors. For (IH, HO-true) combination, there is no CP violation sensitivity at 3$\sigma$ cl.
\item When NH is the true hierarchy, adding data from reactors to LBNE improves CPV sensitivity considerably irrespective of octant. Use of ND helps in improving the sensitivity. For 35 kt FD with ND, 5$\sigma$ discovery is possible irrespective of hierarchy and octant of $\theta_{23}$. For 35 kt FD, fraction of $\delta_{cp}$ to be discovered at 3$\sigma$ and 5$\sigma$ cl with ND is considerably different from that without ND. The reactors combined 35 kt FD (w ND) can discover 0.39 fraction of $\delta_{cp}$ at 5$\sigma$ cl for NH when LO is assumed to be true octant. For IH and LO combination, both LBNE alone and combined analysis     can measure 0.37 fraction of $\delta_{cp}$ at 5$\sigma$ cl.
\item Adding individual reactors can also improve CPV sensitivity of LBNE as seen from fig.3. We have combined Daya Bay and RENO with 35 kt LBNE FD and obsedved that CPV sensitivity increases specially in NH for LO.
\item Adding prior on $\sin^22\theta_{13}$ increases CPV sensitivity when the assumed hierarchy is NH. 
\end{itemize}
To measure CPV phase, it is necessary that LBNE runs in both $\nu$ and $\bar{\nu}$ modes. Since reactor experiments run in $\bar{\nu}$ mode, so, by combining the 5 years $\nu$ data from LBNE with 3 years data from reactors, it is also possible to study CPV. In fig.4, we have combined the 5 years $\nu$ data from 35 kt LBNE FD(w ND) with 3 years data from reactors. It is observed that CP violation is possible to be discovered at more than 3$\sigma$ cl. the fraction of  $\delta_{cp}$  possible to measure at 3$\sigma$ cl is 0.53 for NH irrespective of the octant. In IH mode, the fraction of  $\delta_{cp}$ at 3$\sigma$ cl is found to be 0.37 (0.29) when true octant is the HO (LO).\\
 In fig.2 and fig.3, we have combined both appearance and disappearance channels to study CPV sensitivity. In fig.5, we have explored the possibility of CPV discovery at LBNE with $\nu_\mu \rightarrow \nu_e$ and $\bar{\nu}_\mu \rightarrow \bar{\nu}_e$ channels (appearance) only, as the probability expression for these modes of oscillation contain $\sin\delta$ term which is responsible for CP violation. A comparative study of the three sets of plots- 35 kt FD at LBNE, 35 kt FD at LBNE with reactors and 10 kt FD at LBNE with reactors, have shown that 5$\sigma$ discovery of CPV is possible if 35 kt FD with ND is combined with reactors for that particular channel at LBNE. The sensitivity of the 10 kt plot (with ND and reactors) is not so impressive as that of 35 kt FD (With ND, without reactors) at LBNE.
 
 \section{Conclusions}
 We can draw the following conclusions from the above observations:
 
\begin{itemize}
\item It has been shown in \cite{7}, that LBNE alone would be sensitive to CPV studies. But, CPV discovery potential of LBNE improves significantly in presence of reactor experiments. The difference is prominent specially in NH.
\item CPV discovery potential increases with increase in the mass of the far detector.
\item In presence of ND, CPV sensitivity increases noticeably.
\item Adding reactor experiments in presence of ND with 35 kt FD, makes it possible to discover CPV at 5$\sigma$ cl for both the hierarchies, irrespective of any octant.
 \item Combining Daya Bay, RENO with LBNE individually can also inprove its CPV sensitivity.
\item CPV sensitivity of LBNE in IH mode is higher than that of NH (as can be also seen from fig.1). 
\item In Fig.4, we have presented the possibility of discovering CPV after 5 years run of LBNE in $\nu$ mode when combined with data from reactors. It can provide an early hint of CPV at 3$\sigma$ cl.
\item Only the appearance measurement of LBNE (with ND) can provide hint of CPV discovery at more than 5$\sigma$ cl if combined with reactor experiments.
\end{itemize} 
\begin{figure}[!h]
\begin{centering}
\begin{tabular}{cc}
\includegraphics[width=2.5in]{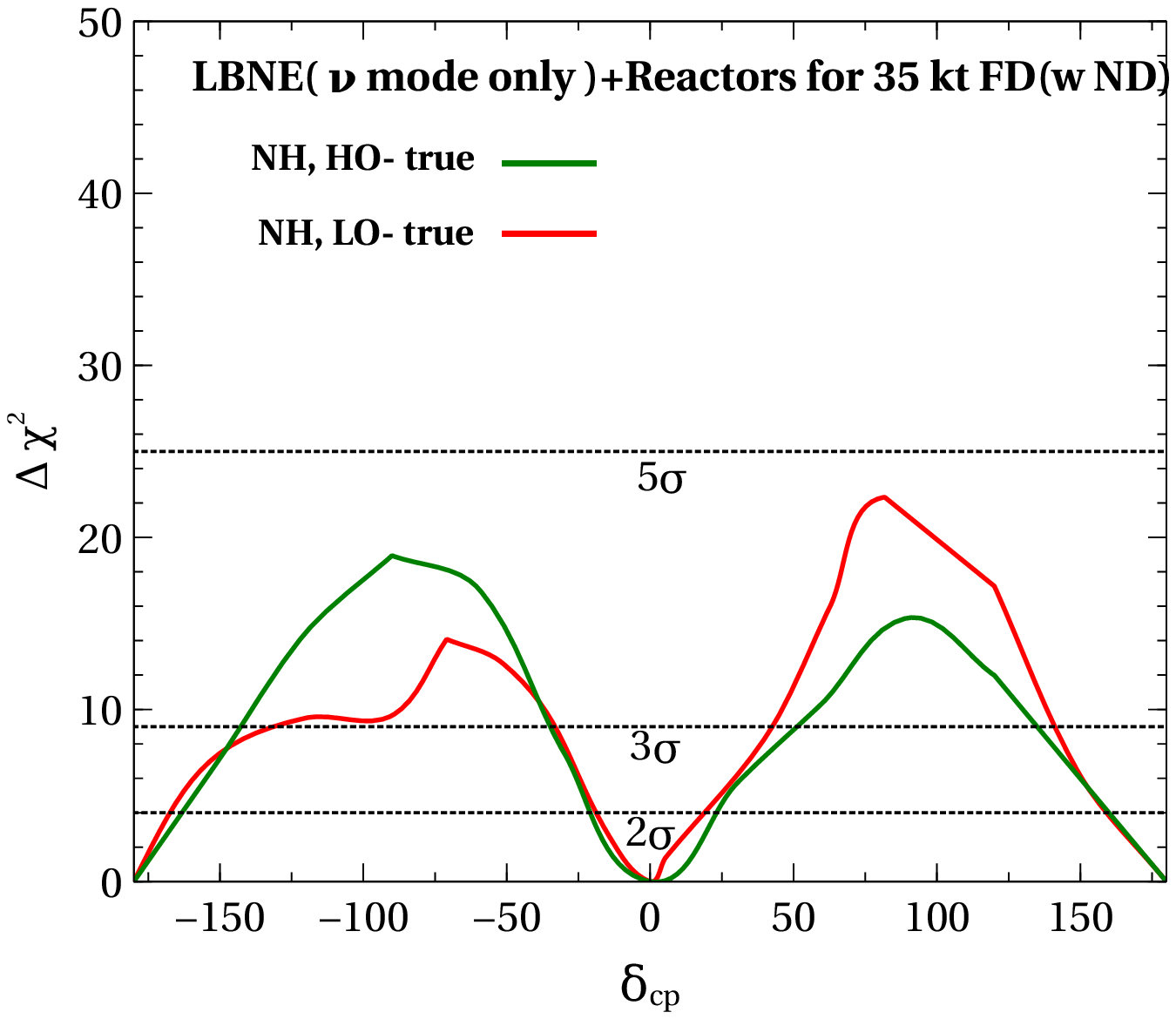}
\includegraphics[width=2.5in]{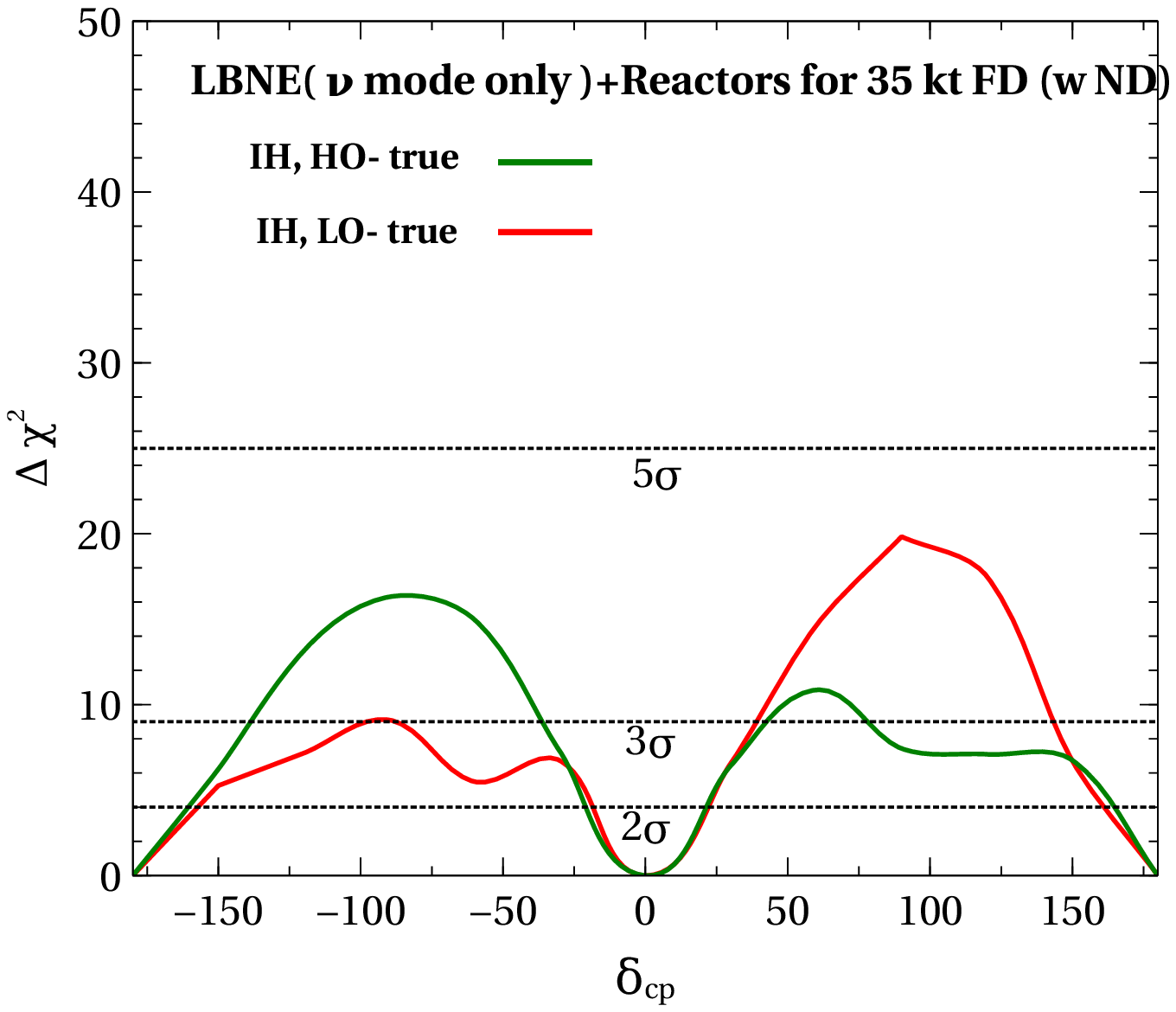}
\end{tabular}
\par\end{centering}
\caption{ CPV sensitivity plots for 5 years LBNE data (in $\nu$ mode) + 3 years data from reactors. Green (red ) solid plot is for true HO (LO) in both the hierarchies. 35 kt FD with ND, when run for 5 years in $\nu$ mode can also discover CP violation if combined with data from reactors. }
\end{figure}

\begin{itemize}
\item Although the appearance channel is responsible for CP violating effect in neutrino sector \cite{7}, combination of both appearance and disappearance channels increases CPV sensitivity. A comparative study of fig.2, fig.3 and fig.5  have confirmed this conclusion. 
\end{itemize}
\section{Acknowledgement}
We thank Pomita Ghoshal for fruitful and constructive suggestions.
\begin{figure}[!h]
\begin{centering}
\begin{tabular}{cc}
\includegraphics[width=2.5in]{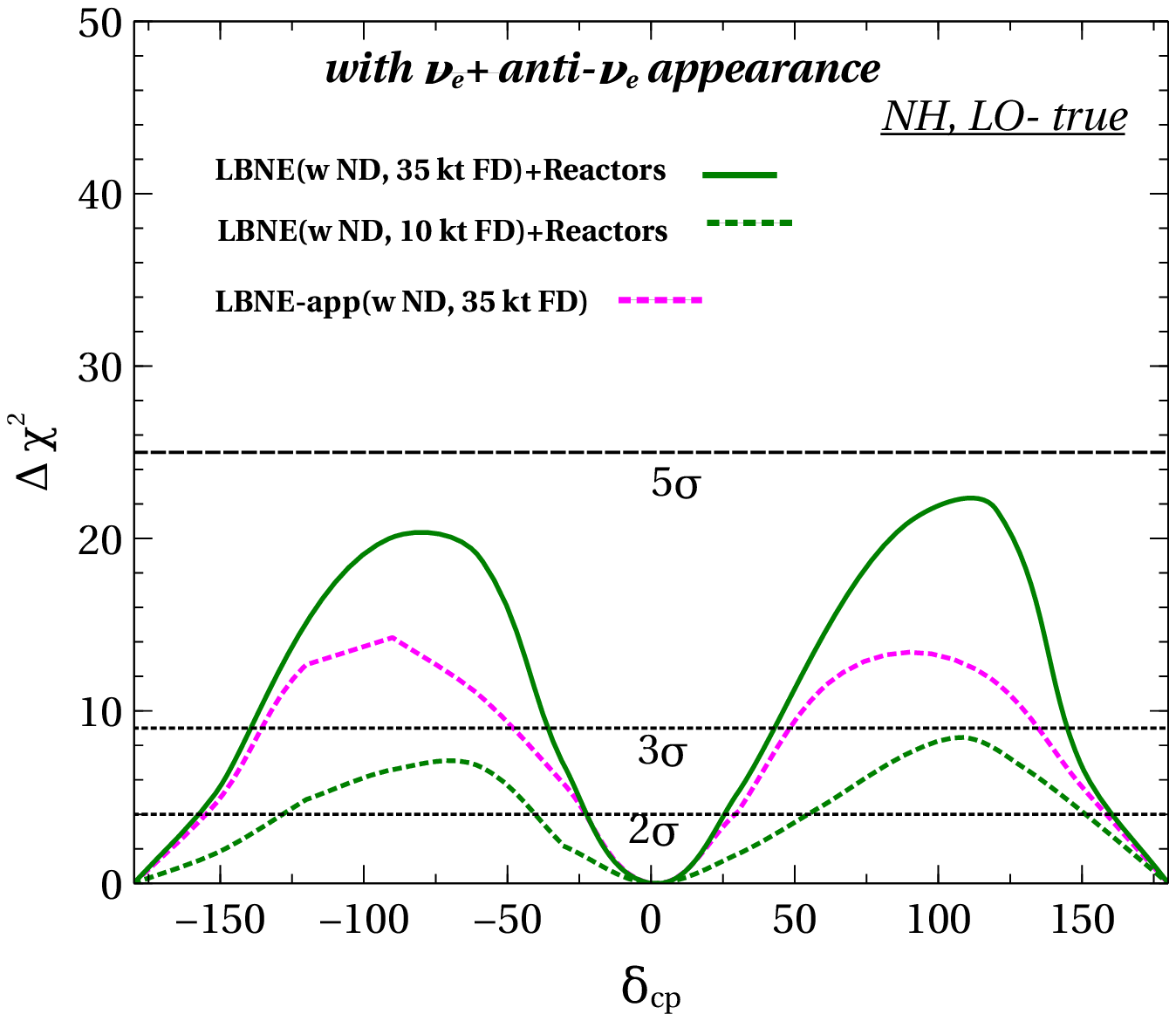}
\includegraphics[width=2.5in]{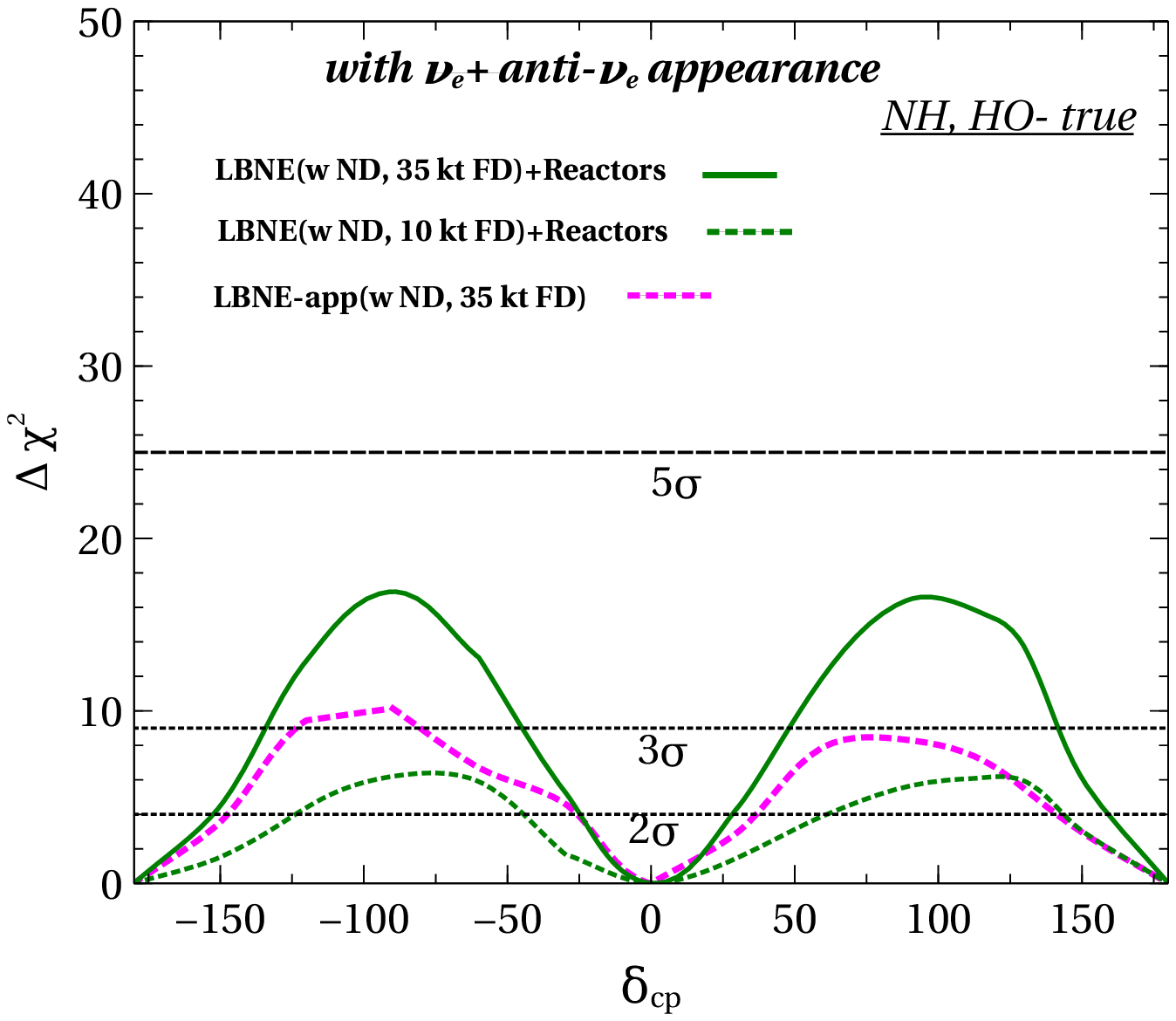}\\
\includegraphics[width=2.5in]{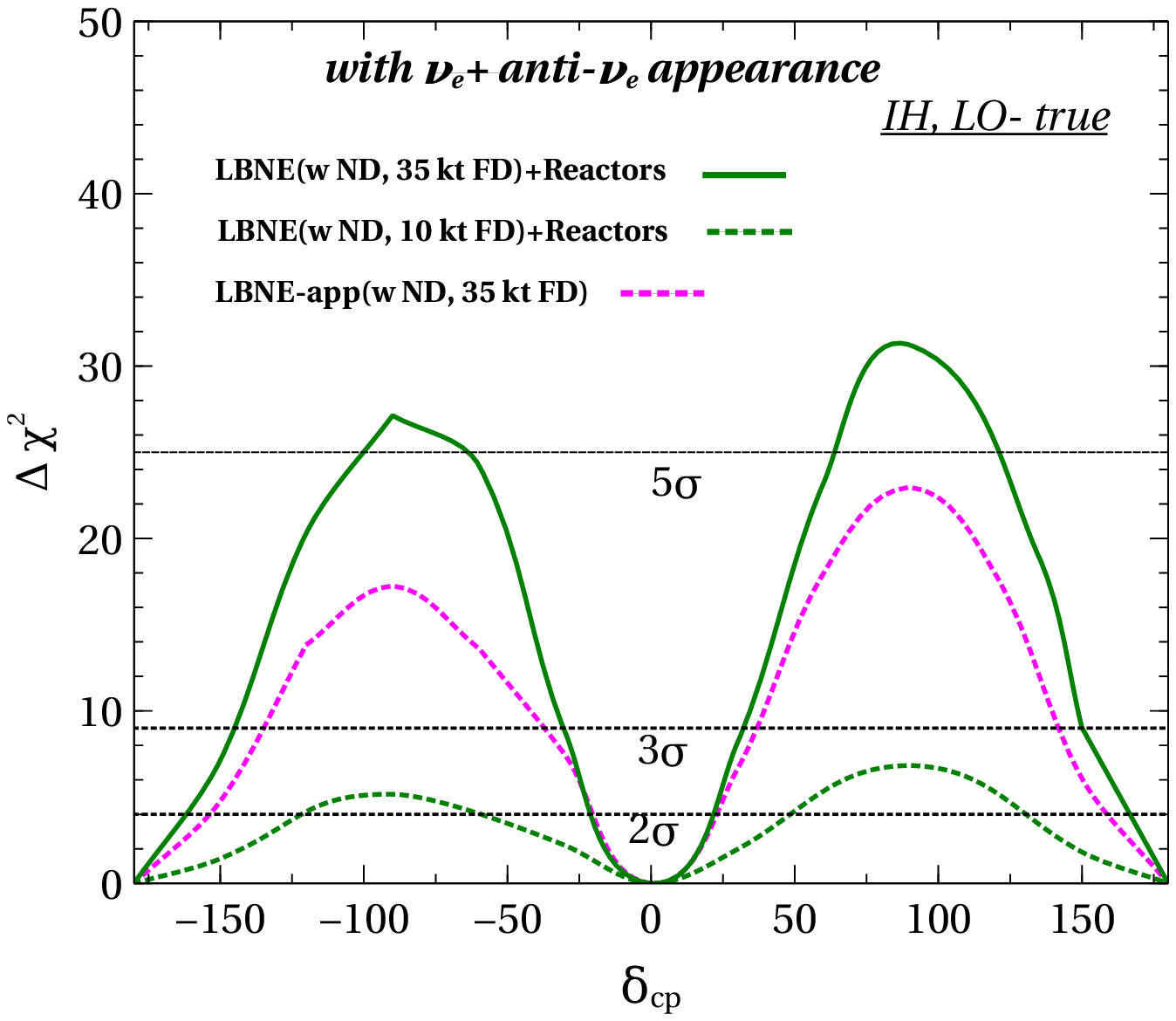}
\includegraphics[width=2.5in]{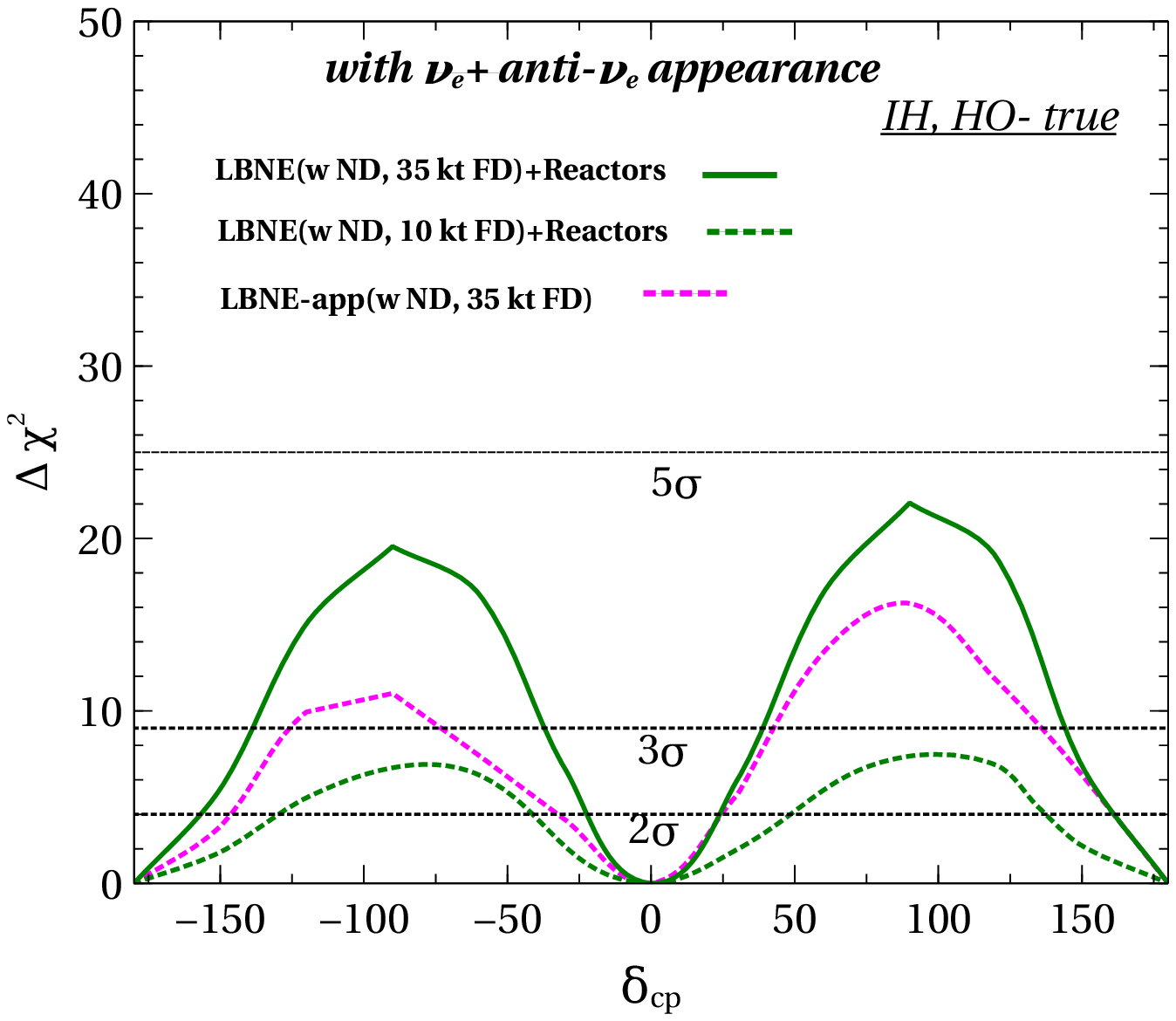}
\end{tabular}
\par\end{centering}
\caption{ CPV sensitivity plots for $\nu_\mu \rightarrow \nu_e$ channel of LBNE in $\nu$ and $\bar{\nu}$ mode. In the upper panel(lower panel), true hierarchy is assumed to be NH(IH). Green solid plot is for reactors combined LBNE for 35 kt FD with ND. Dashed magenta plot is for LBNE with 35 kt FD with ND. Dashed green plot represent 10 kt FD with ND at LBNE, in presence of reactors. In presence of data from reactors (green solid plot), CPV sensitivity increases and CPV is possible to discover at 5$\sigma$ cl. }
\end{figure}

\end{document}